\documentclass[reprint,aps,prd,superscriptaddress,nofootinbib,floatfix,amsfonts,amssymb,amsmath]{revtex4-1}

\usepackage{graphicx,multirow,xspace,comment}
\usepackage[colorlinks=true,citecolor=blue]{hyperref}
\usepackage{breakurl}

\usepackage{amsmath}
\usepackage{amsfonts}
\usepackage{amssymb}

\usepackage{graphicx}

\usepackage{multirow}
\usepackage{rotating}

\usepackage{xcolor}

\usepackage{color}
\definecolor{nicered}{rgb}{0.7,0.1,0.1}
\definecolor{nicegreen}{rgb}{0.1,0.5,.1}
\definecolor{niceblue}{rgb}{0.1,0.2,0.8}

\usepackage{hyperref}
\hypersetup{colorlinks,citecolor=nicegreen,linkcolor=nicered}
\hypersetup{colorlinks=true}

\usepackage{comment}

\usepackage{array}

\usepackage{url}

\begin{document}

\title{Effects of squared four-fermion operators of the \\ Standard Model Effective Field Theory on meson mixing}

\author{Luiz Vale Silva}
\affiliation{Departament de F\'{i}sica Te\`{o}rica, Instituto de F\'{i}sica Corpuscular,
Universitat de Val\`encia -- Consejo Superior de Investigaciones Cient\'{i}ficas,
Parc Cient\'{i}fic, Catedr\'{a}tico Jos\'{e} Beltr\'{a}n 2, E-46980 Paterna, Valencia, Spain}


\begin{abstract}
    The Standard Model Effective Field Theory (SMEFT) is a universal way of parametrizing New Physics (NP) manifesting as new, heavy particle interactions with the Standard Model (SM) degrees of freedom, that respect the SM gauged symmetries. Higher order terms in the NP interactions possibly lead to sizable effects, mandatory for meaningful phenomenological studies, such as contributions to neutral meson mixing, which typically pushes the scale of NP to energy scales much beyond the reach of direct searches in colliders. I discuss the leading-order renormalization of double-insertions of dimension-6 four-fermion operators that change quark flavor by one unit (i.e., $ | \Delta F | = 1 $, $ F = $ strange-, charm-, or bottom-flavor), by dimension-8 operators relevant to meson mixing (i.e., $ | \Delta F | = 2 $) in SMEFT. Then, I consider the phenomenological implications of contributions proportional to large Yukawas, setting bounds on the Wilson coefficients of operators of dimension-6 via the leading logarithmic contributions. Given the underlying interest of SMEFT to encode full-fledged models at low energies, this work stresses the need to consider dimension-8 operators in phenomenological applications of dimension-6 operators of SMEFT.
\end{abstract}



\maketitle


\section{Introduction}

One strategy in searching for signs of New Physics (NP) -- namely, phenomena that cannot be accommodated within the Standard Model (SM) --
is the study of observables that are predicted by the SM to be suppressed, as for instance in the case of flavour changing neutral currents (FCNCs) due to the Glashow-Iliopoulos-Maiani (GIM) mechanism \cite{Glashow:1970gm}.
A different strategy consists in looking for deviations in observables that are precisely predicted, such as the observables that contribute to the extraction of the elements of the Cabibbo-Kobayashi-Maskawa (CKM) matrix \cite{Cabibbo:1963yz,Kobayashi:1973fv} in the SM, among which meson mixing observables, which also fall into the previous category, play an important role \cite{Charles:2015gya,UTfit:2006vpt}.

In the SM, since the latter observables must involve,
compared to the initial and final states
and momenta of external legs, the exchange of much heavier degrees of freedom (e.g., $ W, Z $ bosons), an Effective Field Theory (EFT) provides at low energies a simpler picture of the underlying high-energy dynamics,
in which Wilson coefficients and higher-dimensional operators carry the fingerprints of such heavy particles, a prominent example being Fermi's contact interaction.
Similarly, EFTs can be used to investigate the effects of non-SM new heavy degrees of freedom (e.g., $ W', Z' $), that lead to contact interactions among the SM particles at low enough energies.
The latter EFTs consist of higher-dimensional operators suppressed by some new large scale $ \Lambda_{\rm NP} $, typical of the NP extension, that encode in particular the flavour aspects of the new heavy sector, and their manifestation in observables that are suppressed in the SM or in observables that are precisely predicted can provide clear hints towards the discovery of NP.

The Standard Model Effective Field Theory (SMEFT) consists of the whole set of higher dimensional contact interactions that are consistent with Lorentz and the local symmetries of the SM,
and is particularly useful when a new, weak interacting sector is considered, in which case observational effects are dominated by the first terms in the power series in $ 1 / \Lambda_{\rm NP} $.
In the case of operators of dimension-6, the so-called Warsaw basis \cite{Grzadkowski:2010es} is divided into eight categories, among which we have four-fermions, $ \psi^4 $, that will play a central role in the discussion below, see Tab.~\ref{tab:dimension_6} where we display operators preserving total baryon number.
Explicit on-shell bases for dimension-8 operators have been built \cite{Murphy:2020rsh,Li:2020gnx}, among which one identifies operators involving four fermions, also central to our discussion, see Tab.~\ref{tab:dimension_8} for a subset of them. Redundant operators are discussed in Refs.~\cite{Chala:2021cgt,Ren:2022tvi}.

\begin{table}[]
    \centering
    \renewcommand{\arraystretch}{1.2}
    \begin{tabular}{l}
        $ ( \bar{L} L ) ( \bar{L} L ) $ \\
        \hline
        $ Q^{}_{\ell \ell} = (\bar{\ell} \gamma_\mu \ell) (\bar{\ell} \gamma^\mu \ell) $ \\
        $ Q^{(1)}_{q q} = (\bar{q} \gamma_\mu q) (\bar{q} \gamma^\mu q) $ \\
        $ Q^{(3)}_{q q} = (\bar{q} \gamma_\mu \tau^I q) (\bar{q} \gamma^\mu \tau^I q) $ \\
        $ Q^{(1)}_{\ell q} = (\bar{\ell} \gamma_\mu \ell) (\bar{q} \gamma^\mu q) $ \\
        $ Q^{(3)}_{\ell q} = (\bar{\ell} \gamma_\mu \tau^I \ell) (\bar{q} \gamma^\mu \tau^I q) $ \\
    %
        \phantom{} \\
        $ ( \bar{R} R ) ( \bar{R} R ) $ \\
        \hline
        $ Q^{}_{e e} = (\bar{e} \gamma_\mu e) (\bar{e} \gamma^\mu e) $ \\
        $ Q^{}_{x x} = (\bar{x} \gamma_\mu x) (\bar{x} \gamma^\mu x) $ \\
        $ Q^{}_{e x} = (\bar{e} \gamma_\mu e) (\bar{x} \gamma^\mu x) $ \\
        $ Q^{(1)}_{u d} = (\bar{u} \gamma_\mu u) (\bar{d} \gamma^\mu d) $ \\
        $ Q^{(8)}_{u d} = (\bar{u} \gamma_\mu T^A u) (\bar{d} \gamma^\mu T^A d) $ \\
    %
        \phantom{} \\
        $ ( \bar{L} L ) ( \bar{R} R ) $ \\
        \hline
        $ Q^{}_{\ell e} = (\bar{\ell} \gamma_\mu \ell) (\bar{e} \gamma^\mu e) $ \\
        $ Q^{}_{\ell x} = (\bar{\ell} \gamma_\mu \ell) (\bar{x} \gamma^\mu x) $ \\
        $ Q^{}_{q e} = (\bar{q} \gamma_\mu q) (\bar{e} \gamma^\mu e) $ \\
        $ Q^{(1)}_{q x} = (\bar{q} \gamma_\mu q) (\bar{x} \gamma^\mu x) $ \\
        $ Q^{(8)}_{q x} = (\bar{q} \gamma_\mu T^A q) (\bar{x} \gamma^\mu T^A x) $ \\
    %
        \phantom{} \\
        $ ( \bar{L} R ) ( \bar{R} L ) $ + h.c. \\
        \hline
        $ Q^{}_{\ell e d q} = (\bar{\ell}^m e) (\bar{d} q_m) $ \\
    %
        \phantom{} \\
        $ ( \bar{L} R ) ( \bar{L} R ) $ + h.c. \\
        \hline
        $ Q^{(1)}_{q u q d} = (\bar{q}^m u) \epsilon_{mn} (\bar{q}^n d) $ \\
        $ Q^{(8)}_{q u q d} = (\bar{q}^m T^A u) \epsilon_{mn} (\bar{q}^n T^A d) $ \\
        $ Q^{(1)}_{\ell e q u} = (\bar{\ell}^m e) \epsilon_{mn} (\bar{q}^n u) $ \\
        $ Q^{(3)}_{\ell e q u} = (\bar{\ell}^m \sigma_{\mu \nu} e) \epsilon_{mn} (\bar{q}^n \sigma^{\mu \nu} u) $ \\
    \end{tabular}
    \caption{Four-fermion operators of the so-called Warsaw basis, where $ x = u, d $; $q$ ($\ell$) are weak-isospin doublet quarks (leptons), and $u, d$ ($e$) are weak-isospin singlet quarks (leptons). Flavour or generation indices are omitted; when indicated in the text (e.g., as in the Wilson coefficient $ C_{\ell e d q; f i j k} $) they correspond to the fields above in that same ordering (i.e., $ (\bar{\ell}^m_f e_i) (\bar{d}_j q_{m; k}) $).}
    \label{tab:dimension_6}
\end{table}

\begin{table}[]
    \centering
    \renewcommand{\arraystretch}{1.2}
	\begin{tabular}{l}
		$ (\bar{L} L) (\bar{L} L) H^2 $ \\
		\hline
		$ Q^{(1)}_{\ell^4 H^2} = (\bar{\ell} \gamma_{\mu} \ell) (\bar{\ell} \gamma^{\mu} \ell) (H^\dagger H) $ \\
		$ Q^{(2)}_{\ell^4 H^2} = (\bar{\ell} \gamma_{\mu} \ell) (\bar{\ell} \gamma^{\mu} \tau^I \ell) (H^\dagger \tau^I H) $ \\
		$ Q^{(1)}_{q^4 H^2} = (\bar{q} \gamma_{\mu} q) (\bar{q} \gamma^{\mu} q) (H^\dagger H) $ \\
		$ Q^{(2)}_{q^4 H^2} = (\bar{q} \gamma_{\mu} q) (\bar{q} \gamma^{\mu} \tau^I q) (H^\dagger \tau^I H) $ \\
	%
	    \phantom{} \\
		$ (\bar{R} R) (\bar{R} R) H^2 $ \\
		\hline
		$ Q^{}_{e^4 H^2} = (\bar{e} \gamma_{\mu} e) (\bar{e} \gamma^{\mu} e) (H^\dagger H) $ \\
		$ Q^{}_{x^4 H^2} = (\bar{x} \gamma_{\mu} x) (\bar{x} \gamma^{\mu} x) (H^\dagger H) $ \\
	%
	    \phantom{} \\
		$ (\bar{L} R) (\bar{L} R) H^2 $ \\
		\hline
		$ Q^{(3)}_{\ell^2 e^2 H^2} = (\bar{\ell} e H) (\bar{\ell} e H) $ \\
		$ Q^{(5)}_{q^2 u^2 H^2} = (\bar{q} u \tilde{H}) (\bar{q} u \tilde{H}) $ \\
		$ Q^{(6)}_{q^2 u^2 H^2} = (\bar{q} T^A u \tilde{H}) (\bar{q} T^A u \tilde{H}) $ \\
		$ Q^{(5)}_{q^2 d^2 H^2} = (\bar{q} d H) (\bar{q} d H) $ \\
		$ Q^{(6)}_{q^2 d^2 H^2} = (\bar{q} T^A d H) (\bar{q} T^A d H) $ \\
	\end{tabular}
	\caption{Dimension-8 operators relevant to our discussion, where $ x = u, d $. A complete and minimal basis is found in Ref.~\cite{Murphy:2020rsh}. 
	Flavour or generation indices are omitted, see caption of Tab.~\ref{tab:dimension_6}.}\label{tab:dimension_8}
\end{table}

Operators of dimension-8 may have important phenomenological effects, and started to be discussed more systematically in various contexts:
EW precision tests and Higgs measurements \cite{Chala:2018ari,Hays:2018zze,Hays:2020scx,Corbett:2020bqv,Corbett:2021eux,Trott:2021vqa,Corbett:2021cil,Martin:2021cvs,Dawson:2021xei,Corbett:2023qtg},
collider signals \cite{Alioli:2020kez,Boughezal:2021tih,Kim:2022amu,Li:2022rag,Dawson:2022cmu,Boughezal:2022nof,Degrande:2023iob,Martin:2023tvi},
lepton flavour violation \cite{Ardu:2021koz,Ardu:2022pzk},
gluonic couplings of leptons \cite{Potter:2012yv,Hayreter:2013vna,Cai:2018cog,Cheung:2019bkw},
electron Electric Dipole Moment \cite{Panico:2018hal},
different consequences of causality, analyticity and unitarity requirements \cite{Remmen:2019cyz,Remmen:2020vts,Gu:2020ldn,Bonnefoy:2020yee,Chala:2023jyx,Chen:2023bhu,Chala:2023xjy},
triple neutral gauge couplings \cite{Ellis:2019zex,Ellis:2020ljj,Gu:2020ldn,Ellis:2022zdw},
matching \cite{Hamoudou:2022tdn} and UV completions \cite{Dawson:2022cmu,Banerjee:2022thk,Ellis:2023zim,Banerjee:2023iiv,Li:2023pfw}.
See Refs.~\cite{AccettulliHuber:2021uoa,DasBakshi:2022mwk,Assi:2023zid,Chala:2023xjy} for discussions about the renormalization of single-insertions of operators of dimension-8.

In the case of NP effective operators involving fermion fields,
effects that change flavour by one unit naturally lead to NP effects that change flavour by two units, which in the quark sector are efficiently probed by meson mixing.
This is going to be the main interest here, namely, the leading effect of double-insertions of dimension-6 encoded in dimension-8 operators. To spell out, the focus is on the renormalization of such double-insertions, and the phenomenological limits that can thus be set on the Wilson coefficients of dimension-6 operators. Focusing on the leading order, we will thus overlook a series of issues relevant at higher orders.
Although here we focus on meson mixing, a similar discussion would hold for rare decays, which are loop suppressed in the SM; e.g., see Ref.~\cite{Buchalla:1993wq} in the cases of rare kaon decays.

The leading-order calculation discussed here gives a first quantitative assessment of the size of contributions to meson mixing of double-insertions of dimension-6 operators in SMEFT, and higher-order effects are delegated to future work.
In this respect, the phenomenological importance of meson mixing observables has triggered higher-order calculations in some specific extensions of the SM due to potentially large perturbative QCD corrections, including: two-Higgs-doublet model \cite{Urban:1997gw}, supersymmetry \cite{Ciuchini:1998ix}, left-right model \cite{Bernard:2015boz}, leptoquark model \cite{Crivellin:2021lix}.

Double-insertions of higher-dimensional operators have been discussed in the literature in relation to other problems.
Double-insertion of operators of dimension five have been discussed for instance in Ref.~\cite{Davidson:2018zuo}, that considers the lepton number violating dimension-5 Weinberg operator of SMEFT, and Ref.~\cite{Jenkins:2017dyc}, in the case of a low-energy EFT respecting EM and QCD local symmetries.
The renormalization of double-insertions of operators of dimension-5 and -6 by operators of dimension up to 7 in SMEFT has been discussed in Ref.~\cite{Chala:2021juk}. The renormalization of double-insertions of bosonic operators of dimension-6 has been discussed in Ref.~\cite{Chala:2021pll}, and double-insertions of fermionic operators mediating lepton flavour violation have recently been discussed in Ref.~\cite{Ardu:2022pzk}.
Double-insertions for gluon fusion in collider processes are discussed in Refs.~\cite{Heinrich:2022idm,Asteriadis:2022ras}.
We do not consider in this paper the extension of SMEFT to include new, light degrees of freedom, such as right-handed neutrinos, that can carry a Majorana mass term; see e.g. Ref.~\cite{Liao:2016qyd} for a basis of operators up to dimension-7.

This paper is organized as follows: in Sec.~\ref{sec:effective_operators} we briefly discuss the basis of operators needed in SM calculations of meson mixing at the leading order, and extend the discussion to SMEFT; in Sec.~\ref{sec:large_Yukawa_contributions} we specify our discussion in SMEFT to cases proportional to Yukawa couplings relatively large compared to the Yukawas of external fermion fields; in Sec.~\ref{sec:phenomenology_tops_loops} we discuss phenomenological implications when top-quarks in loops are considered, and then conclude.
Apps.~\ref{sec:AD_tensors} and \ref{sec:sensitivity_NP} contain respectively the explicit expressions of the anomalous dimensions appearing in the RG equations, and a discussion of the sensitivity of meson mixing observables to NP.

\section{Effective operators in meson mixing at the Leading-Log approximation}\label{sec:effective_operators}

\subsection{Standard Model}

At low enough energies, heavy degrees of freedom are integrated out and their dynamics is encoded in the Wilson coefficients of higher-dimensional operators.
One illustration of the use of an EFT is provided by meson mixing in the kaon sector in the SM \cite{Gilman:1982ap,Herrlich:1993yv,Herrlich:1996vf}, which proceeds via box diagrams at the leading order.
Different internal flavours of the same type (here, up-type) can be combined as a result of the GIM mechanism,
which suppresses SM contributions to meson mixing.
There are three sets of contributions that are qualitatively very different, and quantitatively important, according to the elements of the CKM matrix: boxes involving (I) top- and up-,\footnote{Case (I) is the contribution that is largely dominant in neutral-$B_{(s)}$ meson mixing in the SM.} (II) charm- and up-, (III) charm-, top- and up-quarks.\footnote{The same qualitative discussion holds for the different gathering of contributions considered in Ref.~\cite{Brod:2019rzc}.}
At the matching scale $ \mu_{EW} $ (where $ W, Z, H, t $ particles are integrated out and the first EFT is built from the full SM), case (I) is reproduced in the EFT by dimension-6 operators that change flavour by two units ($ | \Delta F | = 2 $), and at the leading order cases (II) and (III) by dimension-6 operators that change flavour by one unit ($ | \Delta F | = 1 $).
Higher orders from strong \cite{Herrlich:1993yv,Herrlich:1996vf,Brod:2010mj,Brod:2011ty,Brod:2019rzc} and electroweak \cite{Brod:2021qvc,Brod:2022har} interactions introduce new operators;
the basis has also to be extended to account for $1/m_c^2$ corrections, see Ref.~\cite{Ciuchini:2021zgf} for a recent reference, and Refs.~\cite{Cata:2003mn,Cata:2004ti}.
A further suppression due to the GIM mechanism is the absence of logarithmic contributions in case (II);
in a consequent way,
it makes double-insertions of dimension-6 operators finite in this case, i.e., the latter do not require renormalization by dimension-8 operators.
GIM does not operate in the same way in case (III), for which the main contribution is given by a large logarithm;
consequently,
GIM does not eliminate the divergence in double-insertions of dimension-6 operators, and case (III) requires renormalization by dimension-8 operators.

To describe the resulting mixing of operators quantitatively, one must determine the anomalous dimension tensor $ \gamma_{ij, n} $: given a set of Green's functions with two insertions of dimension-6 operators (indexed by $ i, j $) one calculates the counter-terms proportional to dimension-8 operators (indexed by $ n $), needed to renormalize the divergences resulting from the double-insertions. Large logarithms are resummed via the Renormalization Group (RG) evolution, see App.~\ref{sec:AD_tensors}:

\begin{equation}\label{eq:RGEs_dim8_dim6}
	\mu \frac{d}{d \mu} C^{(8)}_n (\mu) = \Sigma_{i, j} C^{(6)}_i (\mu) C^{(6)}_j (\mu) \gamma_{ij, n} + \Sigma_m C^{(8)}_m (\mu) \tilde{\gamma}_{m n} \,,
\end{equation}

\noindent
where the superscripts of the Wilson coefficients give the dimension of the corresponding operator.
Solving these RG equations, the term proportional to two dimension-6 Wilson coefficients carries the logarithm $ \log ( \mu_{low} / \mu_{EW} ) $ for some $ \mu_{low} \ll \mu_{EW} $, consistently reproducing the logarithmic enhancement of case (III) above.
The values of the dimension-8 Wilson coefficients $ C^{(8)}_n ( \mu_{EW} ) $ are sub-leading,
and the calculation of the anomalous dimension matrix $ \tilde{\gamma}_{m n} $ is not required at the leading order.

\begin{figure}[t]
    \centering
    \includegraphics[scale=0.3]{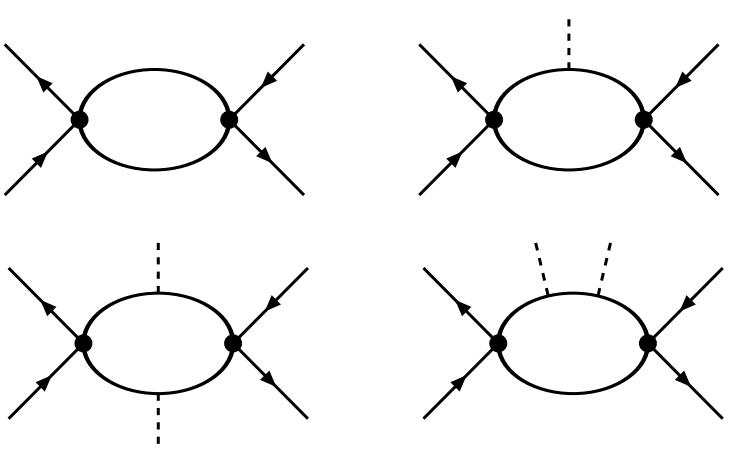}
    \caption{1PI diagrams involving double-insertions and four external fermion legs, represented by solid lines. The Higgs scalar is represented by a dashed line. Gauge bosons can be attached to the internal lines in all possible ways.}
    \label{fig:diagrams_2}
\end{figure}

\subsection{Beyond the Standard Model}

An analogous picture can be drawn in SMEFT.
We consider a case analog to case (III) above.
First of all,
we consider that the underlying NP sector does not generate tree or one-loop contributions to dimension-6 $ | \Delta F | = 2 $ operators, and tree contributions to $ | \Delta F | = 2 $ operators of higher dimension, or at least that these contributions are highly suppressed.\footnote{For instance, in typical left-right models without the addition of new fermions beyond right-handed neutrinos, the masses of the extended (neutral) scalar sector suppress their tree-level contributions to meson mixing, see e.g. Refs.~\cite{Bernard:2015boz,Endo:2018gdn}. In models with leptoquarks, there are no possible tree-level contributions to meson mixing, and integrating out heavy leptoquarks would result in the double-insertions under discussion.}
Then,
we
consider that a possible GIM-like mechanism in the NP sector does not eliminate the need to renormalize double-insertions of $ | \Delta F | = 1 $ dimension-6 operators (whose Wilson coefficients are taken to be non-zero, and uncorrelated a priori), i.e., large logarithms are present, and assumed to be dominant.
Under these assumptions, the leading contribution to meson mixing is captured by double-insertions of dimension-6 operators, that require renormalization by dimension-8 operators. Solving an equation analogous to Eq.~\eqref{eq:RGEs_dim8_dim6} valid in the context of SMEFT results then, when $ \Lambda_{\rm NP} \gg \mu_{EW} $, in the large logarithm $ \log ( \mu_{EW} / \Lambda_{\rm NP} ) $.

For simplicity, we focus on double-insertions of the same operator, with the same flavour content.
Concrete extensions of the SM will typically involve more than one effective operator and a richer flavour structure, and the leading-log contributions therein could be evaluated via a calculation similar to the one described below.
Furthermore, we focus on double-insertions of the full set of dimension-6 four-fermion operators of SMEFT, displayed in Tab.~\ref{tab:dimension_6}.
Our starting point is the Warsaw basis of operators of dimension-6, and we will not discuss the matching of a particular model of renormalizable interactions onto that basis.

\section{Contributions proportional to large Yukawa couplings}\label{sec:large_Yukawa_contributions}
For phenomenological reasons, we focus on cases proportional to relatively large Yukawa couplings compared to the Yukawas of the external fields (which will lead to sizable contributions as we will see below). For instance, we focus on the contributions to kaon-meson mixing that can involve the Yukawas of charm-, bottom- and top-quarks, as well as of tau-leptons.
%

Fig.~\ref{fig:diagrams_2} shows 1PI Feynman diagrams for double-insertions of four-fermion operators that can lead to contributions to meson mixing.
Operators of the schematic structure $ \psi^4 H^2 $ are an obvious candidate to renormalize the divergences of the diagrams in Fig.~\ref{fig:diagrams_2}.
To full generality, other dimension-8 operators also show up in the renormalization programme, schematically: $ \psi^4 H D $, $ \psi^4 D^2 $, and $ \psi^4 X $, see the basis in Ref.~\cite{Murphy:2020rsh}. However, their contributions to meson mixing are proportional to one or two powers of the external fermion masses (of the same order of the external momentum scale), and we will thus neglect their contributions.

The top two diagrams in Fig.~\ref{fig:diagrams_2} introduce in general the need for counter-terms of the structure $ \psi^4 H^2 $, as seen from the equations of motion (EOMs) of fermion fields resulting from the SM Lagrangian.\footnote{\label{ft:shifts_fields}See Refs.~\cite{Passarino:2016saj,Passarino:2019yjx,Criado:2018sdb} and references therein for a discussion in terms of field redefinitions. The use of EOMs is sufficient at the leading order considered here.}
However, these are also suppressed by the external fermion masses.
Therefore, only the bottom two diagrams in Fig.~\ref{fig:diagrams_2} can lead to contributions proportional to large Yukawa couplings.

\begin{figure}[t]
    \centering
    \includegraphics[scale=0.3]{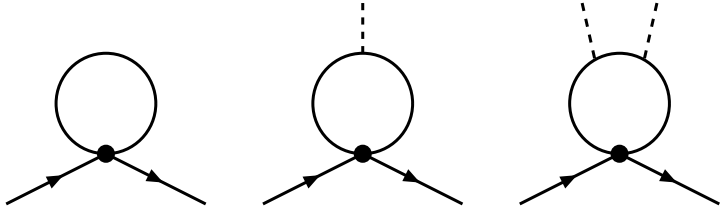}
    \caption{1PI diagrams involving single-insertions and two external fermion legs, represented by solid lines. The Higgs scalar is represented by a dashed line. Gauge bosons can be attached to the internal lines in all possible ways.}
    \label{fig:diagrams_1}
\end{figure}

In principle, 1PI single-insertions of four-fermion operators lead to new contributions, see Fig.~\ref{fig:diagrams_1}. The reason is that higher-dimensional operators change the EOMs of the SM fields \cite{Barzinji:2018xvu}.\footnotemark[4]
However, EOMs of the scalar field and field strength tensors cannot change in presence of dimension-6 four-fermion operators at tree level, and when using the EOMs of fermion fields we end up with contributions to meson mixing suppressed by the external fermion masses, as in the previous two paragraphs.

Basic expressions and explicit anomalous dimensions describing the mixing of double-insertions into operators of the structure $ \psi^4 H^2 $ are collected in App.~\ref{sec:AD_tensors}.
Other dimension-8 operators
would be needed in phenomenological studies when discussing, e.g., tau-lepton loops in the context of charm or bottom physics.

\section{Phenomenology of top-quarks in loops}\label{sec:phenomenology_tops_loops}

We focus our phenomenological discussion on contributions to meson mixing in which both internal fermions running in the loop are top-quarks.
The resulting effect is then proportional to two powers of the Yukawa of the top-quark.
The scope of SMEFT operators is the following, see Tab.~\ref{tab:dimension_6}:

\begin{equation}\label{eq:scope_dimension_6_top_loops}
    Q^{(1, 8)}_{q u q d} \,, \quad Q^{(1, 8)}_{u d} \,, \quad Q^{(1, 8)}_{q u} \,, \quad Q^{(1, 8)}_{q d} \,, \quad Q^{(1, 3)}_{q q} \,.
\end{equation}
Since here the internal flavour is a top, below the EW scale the relevant operators are:

\begin{eqnarray}\label{eq:low_energy_basis}
	&& O^{q_1 q_2}_1 = (\bar{q}_1^\alpha \gamma^\mu L q_2^\alpha) \, (\bar{q}_1^\beta \gamma_\mu L q_2^\beta) \,, \\
	&& O^{q_1 q_2}_2 = (\bar{q}_1^\alpha L q_2^\alpha) \, (\bar{q}_1^\beta L q_2^\beta) \,, \nonumber\\
	&& O^{q_1 q_2}_3 = (\bar{q}_1^\alpha L q_2^\beta) \, (\bar{q}_1^\beta L q_2^\alpha) \,, \nonumber\\
	&& O^{q_1 q_2}_4 = (\bar{q}_1^\alpha L q_2^\alpha) \, (\bar{q}_1^\beta R q_2^\beta) \,, \nonumber\\
	&& O^{q_1 q_2}_5 = (\bar{q}_1^\alpha L q_2^\beta) \, (\bar{q}_1^\beta R q_2^\alpha) \,, \nonumber\\
	&& \tilde{O}^{q_1 q_2}_1 = (\bar{q}_1^\alpha \gamma^\mu R q_2^\alpha) \, (\bar{q}_1^\beta \gamma_\mu R q_2^\beta) \,, \nonumber\\
	&& \tilde{O}^{q_1 q_2}_2 = (\bar{q}_1^\alpha R q_2^\alpha) \, (\bar{q}_1^\beta R q_2^\beta) \,, \nonumber\\
	&& \tilde{O}^{q_1 q_2}_3 = (\bar{q}_1^\alpha R q_2^\beta) \, (\bar{q}_1^\beta R q_2^\alpha) \,, \nonumber
\end{eqnarray}
where $ L $ ($ R $) are left- (right-) handed projectors, $ \alpha, \beta $ are color indices, and flavours are $ q_1, q_2 = b, c, s, d, u $.
Their hadronic matrix elements are calculated for instance in Refs.~\cite{Carrasco:2015pra,Dowdall:2019bea}, which are chirally enhanced in the case of kaons.

\begin{table*}
    \tabcolsep 4pt
	\centering
	\renewcommand{\arraystretch}{1.6}
	{\small
	\begin{tabular}{|>{\centering}m{9mm}m{18mm}|>{\centering\arraybackslash}m{27mm}|>{\centering\arraybackslash}m{27mm}|>{\centering\arraybackslash}m{35mm}|}
	\hline
	WC & flavours & $ B $ ($ f = 3 $, $ i = 1 $) & $ B_s $ ($ f = 3 $, $ i = 2 $) & $ K $ ($ f = 2 $, $ i = 1 $) \\
	\hline
	\hline
	\multirow{2}{*}{ $C^{(1)}_{quqd}$ } & $ 3 3 fi \,, (3 3 if)^\ast $ & $ | C^2 | < (7 \, \text{TeV})^{-4} $ & $ | C^2 | < (4 \, \text{TeV})^{-4} $ & $ | \text{Im} \{ C^2 \} | < (70 \, \text{TeV})^{-4} $ \\ 
	\cline{2-5}
	 & $ f 3 3 i \,, (i 3 3 f)^\ast $ & $ | C^2 | < (5 \, \text{TeV})^{-4} $ & $ | C^2 | < (3 \, \text{TeV})^{-4} $ & $ | \text{Im} \{ C^2 \} | < (50 \, \text{TeV})^{-4} $ \\ 
	\hline
	\multirow{2}{*}{ $C^{(8)}_{quqd}$ } & $ 3 3 fi \,, (3 3 if)^\ast $ & $ | C^2 | < (3 \, \text{TeV})^{-4} $ & $ | C^2 | < (2 \, \text{TeV})^{-4} $ & $ | \text{Im} \{ C^2 \} | < (30 \, \text{TeV})^{-4} $ \\ 
	\cline{2-5}
	 & $ f 3 3 i \,, (i 3 3 f)^\ast $ & $ | C^2 | < (3 \, \text{TeV})^{-4} $ & $ | C^2 | < (2 \, \text{TeV})^{-4} $ & $ | \text{Im} \{ C^2 \} | < (30 \, \text{TeV})^{-4} $ \\ 
	\hline
	\end{tabular}
	}
	\caption{Estimated bounds on the Wilson coefficients of the operators in Eq.~\eqref{eq:scope_dimension_6_top_loops}. The first column indicates the Wilson Coefficient (WC) $C$ being probed at the scale $ \Lambda_{\rm NP} $, together with its flavour indices (bounds correspond to the combination of the two cases provided, with no interference present). The three remaining columns give estimated bounds accordingly to the meson system.
    The flavour indices with an asterisk correspond to the WC of the operator that has been complex conjugated; in this way we indicate both cases in which the $f$ flavour comes from a weak-isospin doublet (first cases) or a singlet (second cases).
	}\label{tab:summary_1}
\end{table*}

Finally, one sets constraints on the NP Wilson coefficients of the operators in Eq.~\eqref{eq:scope_dimension_6_top_loops} based on their contributions to different meson mixing observables, namely: indirect CP violation in the system of kaons, and mass differences in the systems of $B_{(s)}$-mesons. These observables are used in the global fit of Ref.~\cite{CKMfitterGroup} and a good global agreement with the SM is presently obtained. Although the relevant RGEs in SMEFT are provided in App.~\ref{sec:AD_tensors}, we do not discuss the mass difference in the system of kaons nor charm-meson mixing, due to the underlying difficulty in having precise SM predictions in these cases in consequence of long-distance effects. Similarly, although our focus is on meson mixing, we also provide in App.~\ref{sec:AD_tensors} the RGEs relevant for muonium oscillation, $ \mu^+ e^- \leftrightarrow \mu^- e^+ $ (for a recent reference on this topic see Ref.~\cite{Conlin:2020veq}). We do not include the latter in our discussion due to the limited sensitivity to the NP scale resulting from the generated $ 1 / \Lambda_{\rm NP}^4 $ effect.
We exploit the bounds provided in Refs.~\cite{Charles:2013aka,Charles:2020dfl} on generic NP contributions, see App.~\ref{sec:sensitivity_NP}.
Despite being an effect proportional to a dimension-8 operator and being generated at one-loop, given the precision with which these observables are known, one reaches a sensitivity to multi-TeV NP effects, see Tabs.~\ref{tab:summary_1} and \ref{tab:summary_2}. We work in the basis in which down-type flavours are mass eigenstates (phenomenological results for down-type meson-mixing are more straightforwardly assessed in this basis).
A certain number of comments is in order:

\begin{itemize}
    \item Sub-leading effects above the EW scale may be numerically relevant if the leading logarithmic term is not largely dominant, as in any similar study. Their determination, however, is beyond the scope of this work.
    Among the possible effects showing up at higher order, we mention single-insertions of $ | \Delta F | = 2 $ dimension-8 operators and their renormalization by $ | \Delta F | = 2 $ dimension-6 operators. Higher orders introduce the need for determining the Wilson coefficients $ C^{(8)} $ calculated at the matching scale $ \Lambda_{\rm NP} $.
    \item
    The fit \cite{Charles:2013aka,Charles:2020dfl} is done under the assumption that NP is only present in contact interactions that change flavour by two units, while here we analyse the combined effects of $|\Delta F|=1$ NP operators. In presence of $|\Delta B|=1$ operators $Q^{(1, 3)}_{q q} $, $ Q^{(1, 8)}_{u d} $, $ Q^{(1, 8)}_{q u}$, $ Q^{(1, 8)}_{q d}$ there is in particular NP affecting the interpretation of the extracted values of the unitary triangle angles $ \beta, \beta_s $ (beyond mixing-induced CP violation, that is already taken into account in that fit), namely, tree contributions involving the charm flavour\footnote{See Refs.~\cite{Jager:2017gal,Jager:2019bgk} for a comprehensive discussion of NP operators mediating $b \to c \bar{c} s$.} (suppressed by off-diagonal elements of the CKM matrix), and/or contributions similar to the (gluonic) penguin generated in the SM.\footnote{For example, in the SM penguin pollution in the extraction of $\beta$ from $B^0 \to J/\psi K^0$ is small, and neglected given current experimental uncertainties: in particular, the imaginary part of the top-penguin contribution to the amplitudes scales like $\lambda^4$, and is doubly-Cabibbo suppressed with respect to the tree contribution (independently of the Wolfenstein parameterization), see e.g. Ref.~\cite{Boos:2004xp}.
    However, NP does not have to follow the same suppression; see e.g. Ref.~\cite{Grossman:2002bu} for a discussion of NP effects in the decay. Moreover, one cannot exploit $SU(3)$ relating $B^0 \to J/\psi K^0$ to $B^0 \to J/\psi \pi^0$ to constrain top-penguins affecting the former channel \cite{Ciuchini:2005mg,Faller:2008zc,Jung:2012mp,Barel:2020jvf} since NP operators carrying different flavours are assumed unrelated.} These observables play a central role in setting constraints on the allowed size of NP in $|\Delta B|=2$ \cite{Lenz:2010gu,Lenz:2012az,Lenz:2019lvd}, and for this reason $|\Delta B|=1$ operators are not included in Tab.~\ref{tab:summary_2}. See also Ref.~\cite{Descotes-Genon:2018foz} for a discussion of the effects of dimension-6 operators in the extraction of the elements of the CKM matrix. A re-analysis of the global fit taking into account $|\Delta B|=1$ operators involving top-quarks will be the subject of future work, further motivated by the fact that experimental uncertainties in the extraction of $ \beta $ will be improved. On the other hand, we provide bounds on NP in the kaon sector in Tab.~\ref{tab:summary_2}, which result in the global fit of Ref.~\cite{Charles:2013aka} from $ | \epsilon_K | $.
    \item In the case of the operators shown in Tab.~\ref{tab:summary_2}, single-insertions \cite{Endo:2018gdn,Silvestrini:2018dos,Hurth:2019ula,Aebischer:2020dsw} at one-loop can lead to contributions to meson mixing,
    setting bounds on $ | \text{Im} \{ V_{td} V_{ts}^\ast C \} | $ instead of $ | \text{Im} \{ C^2 \} | $, see Refs.~\cite{Endo:2018gdn,Hurth:2019ula} where finite terms are discussed, that contribute in the matching to an effective theory valid below the EW energy scale.
    Although these single-insertions are suppressed by off-diagonal elements of the CKM matrix, chiral enhancements in cases $ Q^{(1, 8)}_{u d} $ and $ Q^{(1, 8)}_{q d} $ lead to a much better sensitivity to the NP energy scale compared to Tab.~\ref{tab:summary_2}. In cases $ Q^{(1, 8)}_{q u} $, the sensitivity is similar to the one shown in Tab.~\ref{tab:summary_2}.
    The operators $Q^{(1, 3)}_{q q} $ can contribute to charm-meson mixing (with suppression by off-diagonal elements of the CKM matrix) \cite{Silvestrini:2018dos}, which however is challenging to calculate reliably in the SM.
    In the case of the operators shown in Tab.~\ref{tab:summary_1}, contributions from finite terms are suppressed by the mass scale of the external fields, and they are thus negligible.
    \item Other observables can also constrain the same NP effective couplings; e.g., many other operators are radiatively generated through single-insertions, whose anomalous dimensions at one-loop are found in Refs.~\cite{Jenkins:2013zja,Jenkins:2013wua,Alonso:2013hga}. These radiative effects result in contributions to rare semi-leptonic transitions, for instance. However, in particular, $ K \to \pi \bar{\nu} \nu $ rates are presently not known to an experimental accuracy much better than $100\%$ \cite{Zyla:2020zbs}, and the main sensitivity to the NP scale is still achieved by meson mixing.
    On the other hand, in the cases of the operators $ Q^{(8)}_{qu} $ and $ Q^{(3)}_{qq; f 3 3 i} $, although large theoretical uncertainties are involved, a much higher sensitivity to the NP scale is likely to be achieved by the observable $ \varepsilon'/\varepsilon $ (giving the amount of direct CP violation in the system of kaons) from gluonic penguins due to the chiral enhancement involved \cite{Aebischer:2018quc}, similar to the gluonic LR penguin operator generated in the SM.
\end{itemize}

\noindent
We have thus obtained a dominant, or competitive, clean\footnote{I.e., excluding charm-meson mixing, or alternatively working directly in the mass basis for fermions and assuming that NP leads to dimension-6 operators involving the top and not the charm.}
sensitivity to NP effects from double-insertions in the cases shown in Tab.~\ref{tab:summary_1}, and the cases $ Q^{(1)}_{qu; f i 3 3} $, $ Q^{(1)}_{qq; f i 3 3} $ and $ Q^{(3)}_{qq; f i 3 3} $, shown in Tab.~\ref{tab:summary_2}.
%
Even in cases where a higher sensitivity is reached by single-insertions,
double-insertions
carry a different dependence on the dimension-6 Wilson coefficients, and can therefore offer a complementary probe.


Beyond top-quarks in loops, one can also have other internal heavy flavours in the case of kaon-meson mixing, namely, charm- and bottom-quarks, and tau-leptons, which are much heavier than kaons.
Given the dependence on masses lighter than the top, the sensitivity to the NP scale will drop.

\begin{table}
	\centering
	\renewcommand{\arraystretch}{1.6}
	\begin{tabular}{|>{\centering}m{10mm}m{20mm}|>{\centering\arraybackslash}m{45mm}|}
	\hline
	WC & flavours & $ K $ ($ f = 2 $, $ i = 1 $) \\
	\hline
	\hline
	\multirow{1}{*}{ $C^{(1)}_{ud}$ } & $ 3 3 fi $ & $ | \text{Im} \{ C^2 \} | < (30 \, \text{TeV})^{-4} $ \\ 
	\hline
	\multirow{1}{*}{ $C^{(8)}_{ud}$ } & $ 3 3 fi $ & $ | \text{Im} \{ C^2 \} | < (10 \, \text{TeV})^{-4} $ \\ 
	\hline
	\hline
	\multirow{1}{*}{ $C^{(1)}_{qd}$ } & $ 3 3 fi $ & $ | \text{Im} \{ C^2 \} | < (30 \, \text{TeV})^{-4} $ \\ 
	\hline
	\multirow{1}{*}{ $C^{(8)}_{qd}$ } & $ 3 3 fi $ & $ | \text{Im} \{ C^2 \} | < (10 \, \text{TeV})^{-4} $ \\ 
	\hline
	\hline
	\multirow{1}{*}{ $C^{(1)}_{qu}$ } & $ fi 3 3 $ & $ | \text{Im} \{ C^2 \} | < (30 \, \text{TeV})^{-4} $ \\ 
	\hline
	\multirow{1}{*}{ $C^{(8)}_{qu}$ } & $ fi 3 3 $ & $ | \text{Im} \{ C^2 \} | < (10 \, \text{TeV})^{-4} $ \\ 
	\hline
	\hline
	\multirow{1}{*}{ $C^{(1)}_{qq}$ } & $ fi 3 3 = 3 3 fi $ & $ | \text{Im} \{ C^2 \} | < (30 \, \text{TeV})^{-4} $ \\ 
	\hline
	\multirow{2}{*}{ $C^{(3)}_{qq}$ } & $ fi 3 3 = 3 3 fi $ & $ | \text{Im} \{ C^2 \} | < (30 \, \text{TeV})^{-4} $ \\ 
	\cline{2-3}
	 & $ f 3 3 i = 3 i f 3 $ & $ | \text{Im} \{ C^2 \} | < (30 \, \text{TeV})^{-4} $ \\ 
	\hline
	\end{tabular}
	\caption{See caption of Tab.~\ref{tab:summary_1} for comments. Additionally, the contribution of $ Q^{(1)}_{qq; f 3 3 i} $ to double-insertions proportional to the Yukawa of the top-quark squared vanishes at this order.}\label{tab:summary_2}
\end{table}

\section{Conclusions}

I have discussed effects of a generic heavy NP sector that are encoded in higher-dimensional operators.
More exactly, I have calculated the renormalization by dimension-8 operators of double-insertions of dimension-6 operators, where the latter changes flavour number by one unit and the former by two units.
At energy scales much below the characteristic scale of NP, the effects of double-insertions are constrained by meson mixing observables, which receive suppressed contributions from the SM, due to the GIM mechanism, that are precisely predicted in the case of many observables.
The calculation here discussed provides the leading-order contribution to meson mixing in SMEFT when $|\Delta F| = 2$ tree-level effects and dimension-6 $|\Delta F| = 2$ operators generated at one-loop are absent or suppressed, and no GIM-like mechanism in the NP sector operates to eliminate the need for renormalization.

The scope of SMEFT operators considered here extends to four-fermions of different chiralities, and to semi-leptonic operators.
I have focused the phenomenological discussion on tops as the internal flavour in fermionic loops, resulting in contributions proportional to the square of the Yukawa coupling of the top.
Given the level of experimental accuracy reached for meson mixing observables, loop-suppressed double-insertions lead to meaningful and powerful bounds on NP, displayed in Tabs.~\ref{tab:summary_1} and \ref{tab:summary_2}, probing energy scales much beyond the reach of direct searches in colliders.

\section*{Acknowledgements}

I am indebted to Antonio Pich for early involvement in this project and for careful reading the manuscript.
{This work has been supported in part by MCIN/AEI/10.13039/501100011033 Grant No. PID2020-114473GB-I00, by PROMETEO/2021/071 (GV).}
This project has received funding from the European Union’s Horizon 2020 research and innovation programme under the Marie Sklodowska-Curie grant agreement No 101031558.



\appendix

\section{Renormalization and RG equations}\label{sec:AD_tensors}

We consider on-shell renormalization with dimensional regularization ($ D = 4 - 2 \epsilon $) and (modified) minimal subtraction;
Ref.~\cite{Chetyrkin:1997fm} is used to identify UV divergences;
we also employ BMHV scheme to deal with $ \gamma_5 $.

Next, we follow closely the discussion of Ref.~\cite{Herrlich:1996vf}. It will be left implicit that the operators being discussed are the ones of the main text. Up to $ \mathcal{O} (\Lambda_{\rm NP}^{-6}) $ terms, the non-renormalizable part of the Lagrangian is given by (sums over repeated indices are left implicit):

\begin{eqnarray}\label{eq:non_ren_Lag}
	&& \qquad\qquad \mathcal{L}_{\text{eff; non-ren.}} = \\
	&& - \frac{1}{\Lambda_{\rm NP}^2} \left( \mathcal{C}^{(6)}_i Z^{-1}_{ij} + \frac{m^2_H}{\Lambda_{\rm NP}^2} \mathcal{C}^{(8)}_i \hat{Z}^{-1}_{ij} \right) \mu^{\Delta^{(6)}_j} Q^{(6), \text{bare}}_j \nonumber\\
	&& - \frac{1}{\Lambda_{\rm NP}^4} \left( \mathcal{C}^{(6)}_i \mathcal{C}^{(6)}_j Z^{-1}_{ij, n} + \mathcal{C}^{(8)}_m \tilde{Z}^{-1}_{m n} \right) \mu^{\Delta^{(8)}_n} Q^{(8), \text{bare}}_n \nonumber\\
	&& \qquad\qquad\qquad + \ldots \,, \nonumber
\end{eqnarray}
where ``$ (6) $" and ``$ (8) $" are the dimensions of the operators involved, except when otherwise indicated (namely, the dimension-8 operators $ Q^{(6)}_{q^2 x^2 H^2} $, $ x = u, d $, and their Wilson coefficients).
The Wilson coefficients $ \mathcal{C}^{(6)} $ and $ \mathcal{C}^{(8)} $ are dimensionless (we reserve the typesetting $ C^{(6)} $ and $ C^{(8)} $ for the dimensionful ones), and the expressions multiplying the bare operators are the bare Wilson coefficients.
The ellipses denote counter-terms proportional to unphysical operators, that will be omitted hereafter.
Note that on-shell matrix elements of double-insertions of EOM-vanishing operators are in principle non-zero, see e.g. Ref.~\cite{Simma:1993ky}. However, we consider as our starting point in the main text dimension-6 operators of the Warsaw basis, which we do not replace via the use of EOMs.
Other than its non-renormalizable part, $ \mathcal{L}_{\text{eff; non-ren.}} $, the full Lagrangian $ \mathcal{L}_{\text{eff}} $ also includes $ \mathcal{L}_{``\text{SM}"} $, which is the SM Lagrangian together with counter-terms proportional to SM operators due to the presence of NP interactions.

Insertions of higher-dimensional operators leading to Green's functions relevant for meson mixing are indicated as (the ``SM'' part of the Lagrangian is not being explicitly shown):

\begin{eqnarray}
	&& \langle \mathbf{T} \exp \left[ i \int d^D x \mathcal{L}_{\text{eff; non-ren.}} (x) \right] \rangle^{\text{``SM''}}_{| \Delta F | = 2} = \nonumber \\
	&& \qquad -i \langle a^{(6)} + a^{(8)} \rangle^{\text{``SM''}}_{| \Delta F | = 2}
\end{eqnarray}
up to terms $ \mathcal{O} (\Lambda_{\rm NP}^{-6}) $, where

\begin{equation}
	a^{(6)} (x) = \frac{1}{\Lambda_{\rm NP}^2} \Sigma_{i, j} \mathcal{C}^{(6)}_i Z^{-1}_{ij} \mu^{\Delta^{(6)}_j} Q^{(6), \text{bare}}_j (x)
\end{equation}
and

\begin{eqnarray}
	&& a^{(8)} (x) = \frac{m_H^2}{\Lambda_{\rm NP}^4} \Sigma_{i, j} \mathcal{C}^{(8)}_i \hat{Z}^{-1}_{ij} \mu^{\Delta^{(6)}_j} Q^{(6), \text{bare}}_j (x) \nonumber \\
	&& \qquad + \frac{1}{\Lambda_{\rm NP}^4} \Sigma_{m, n} \mathcal{C}^{(8)}_m \tilde{Z}^{-1}_{m n} \mu^{\Delta^{(8)}_n} Q^{(8), \text{bare}}_n (x) \nonumber\\
	&& \qquad\qquad + \frac{1}{\Lambda_{\rm NP}^4} \Sigma_{i, j} \mathcal{C}^{(6)}_i \mathcal{C}^{(6)}_j \mathcal{R}_{i j} (x) \,,
\end{eqnarray}
with

\begin{eqnarray}
	&& \mathcal{R}_{i j} (x) = \Sigma_{i', j'} Z^{-1}_{i i'} Z^{-1}_{j j'} \mu^{\Delta^{(6)}_{i'} + \Delta^{(6)}_{j'}} \mathcal{R}^{\text{bare}}_{i' j'} (x) \nonumber \\
	&& \qquad + \Sigma_n Z^{-1}_{ij, n} \mu^{\Delta^{(8)}_n} Q^{(8), \text{bare}}_n (x)
\end{eqnarray}
and

\begin{eqnarray}
	&& \mathcal{R}^{\text{bare}}_{i' j'} (x) = \frac{-i}{2} \int d^D y \mathbf{T} \left( Q^{(6), \text{bare}}_{i'} (x) Q^{(6), \text{bare}}_{j'} (y) \right. \nonumber \\
	&& \qquad \left. + Q^{(6), \text{bare}}_{j'} (x) Q^{(6), \text{bare}}_{i'} (y) \right) \,.
\end{eqnarray}

In powers of the Yukawa couplings $ y $ (possibly different Yukawa couplings are represented by the same letter $ y $), the renormalization factors are expanded as ($ \delta $ is the Kronecker symbol, with indices omitted):

\begin{eqnarray}
	&& Z^{-1} = \delta + \frac{y^2}{(4 \pi)^2} Z^{-1,(1)} + \ldots \,, \\
	&& Z^{-1,(n)} = \Sigma^n_{r=0} \frac{1}{\epsilon^r} Z^{-1,(n)}_r + \mathcal{O} (\epsilon) \,, \nonumber
\end{eqnarray}
and similarly for $ \tilde{Z}^{-1} $, while $ \hat{Z}^{-1} $ and the tensor $ Z^{-1}_{ij,n} $ have perturbative expansions starting at $ \mathcal{O} (y^2) $.

From the scale independence of the bare coefficients, one gets (cf. Eq.~\eqref{eq:RGEs_dim8_dim6}, given in the context of the SM)

\begin{equation}\label{eq:RGEs_dim8_dim6_SMEFT}
	\mu \frac{d}{d \mu} \mathcal{C}^{(8)}_n (\mu) = \Sigma_{i, j} \mathcal{C}^{(6)}_i (\mu) \mathcal{C}^{(6)}_j (\mu) \gamma_{ij, n} + \Sigma_m \mathcal{C}^{(8)}_m (\mu) \tilde{\gamma}_{m n} \,,
\end{equation}
with in particular

\begin{eqnarray}\label{eq:definition_gamma0}
	&& \gamma_{ij, n} = \frac{y^2}{(4 \pi)^2} \gamma^{(0)}_{ij, n} + \ldots \,, \\
	&& \gamma^{(0)}_{ij, n} = \frac{\left( 2 \, \Delta_{y} + \Delta^{(6)}_i + \Delta^{(6)}_j - \Delta^{(8)}_n \right)}{\epsilon} [Z^{-1,(1)}_1]_{ij, n} \nonumber\\
    && \qquad + \mathcal{O} (\epsilon) \,, \nonumber
\end{eqnarray}
where $ \Delta_{y} = \epsilon $ is the mass dimension of the coupling $ y $, which satisfies the renormalization group equation $ \mu \, d y (\mu) / d \mu = - \Delta_{y} \, y (\mu) + \mathcal{O} (y^2) $. Similarly, $ \Delta^{(6)} = 2 \epsilon $ for four-fermion operators $ \psi^4 $, and $ \Delta^{(8)} = 4 \epsilon $ for operators of the schematic form $ \psi^4 H^2 $.
To achieve the simple form of Eq.~\eqref{eq:RGEs_dim8_dim6_SMEFT}, we have neglected terms $ \propto m^2_H $ indicated in the right-hand side of Eq.~\eqref{eq:non_ren_Lag}, which is justified in the leading order being considered here, where $ \mathcal{C}^{(8)} (\Lambda_{\rm NP}) $ is sub-leading, see the main text, Sec.~\ref{sec:effective_operators}.

The anomalous dimensions $ \gamma^{(0)}_{ij, n} $ are then calculated from the finiteness of the Green's functions introduced above, $ \langle \mathcal{R}_{i j} \rangle^{(0)} $ in particular (``$ (0) $'' indicates the leading order, shown in Figs.~\ref{fig:diagrams_2} and \ref{fig:diagrams_1}, and ``$ [div.] $'' the divergent parts):
\begin{eqnarray}\label{eq:identifying_div}
	&& \frac{(y^{\text{bare}})^2}{(4 \pi)^2} \Sigma_{n} [ Z^{-1,(1)} ]_{ij, n} \langle Q^{(8), \text{bare}}_n \rangle^{(0), \text{SM}}_{| \Delta F | = 2} \;[div.] = \nonumber \\
	&& \qquad - \langle \mathcal{R}^{\text{bare}}_{i j} \rangle^{(0), \text{SM}}_{| \Delta F | = 2} \;[div.] \,.
\end{eqnarray}
Since we focus on the special case in which $ i = j $, $ \gamma^{(0)} $ is a matrix.

Finally, at the leading-log approximation:

\begin{eqnarray}
	&& \mathcal{C}^{(8)}_n (\mu_{EW}) = \frac{y^2 (\mu_{EW})}{(4 \pi)^2} \; \ell n \left( \frac{\mu_{EW}}{\Lambda_{\rm NP}} \right) \nonumber \\
	&& \qquad \times \Sigma_{i, j} \mathcal{C}^{(6)}_i (\Lambda_{\rm NP}) \mathcal{C}^{(6)}_j (\Lambda_{\rm NP}) \gamma^{(0)}_{ij, n} \,.
\end{eqnarray}
The scale $\Lambda_{\rm NP}$ appearing in the logarithm is set to $1$~TeV in the numerical applications in the main text.

\subsection{Explicit expressions for the RG equations}

The RG equations are given in the following (the sum over $ j, k $ indices is left implicit):


\begin{eqnarray}
	&& \qquad\qquad (4 \pi)^2 \mu \frac{d}{d \mu} C^{(3)}_{\ell^2 e^2 H^2; fifi} = \\
	&& G^{\ell e d q}_{a} [lepton] \times S_{a} \times \mathcal{G}^{\ell e d q}_{a} \times (Y_d^\dagger)_{kj}^2 \times (C_{\ell e d q; fijk})^2 \nonumber \\
	&& + G^{\ell e q u (1)}_{a} [lepton] \times S_{a} \times \mathcal{G}^{\ell e q u (1)}_{a} \nonumber \\
	&& \qquad \times (Y_u)_{kj}^2 \times (C^{(1)}_{\ell e q u; fijk})^2 \nonumber \\
	&& + G^{\ell e}_{b} \times S_{b} \times \mathcal{G}^{\ell e}_{b} \times (Y_e^\dagger)_{kj}^2 \times (C_{\ell e; fkji})^2 \nonumber
\end{eqnarray}

\begin{eqnarray}
	&& \qquad\qquad (4 \pi)^2 \mu \frac{d}{d \mu} C^{(5)}_{q^2 d^2 H^2; fifi} = \\
	&& G^{\ell e d q}_{a} [quark] \times S_{a} \times \mathcal{G}^{\ell e d q}_{a} \times (Y_e^\dagger)_{kj}^2 \times (C^\ast_{\ell e d q; kjif})^2 \nonumber \\
	&& + G^{q u q d (1)}_{a} \times S_{a} \times \mathcal{G}^{q u q d (1)}_{a} \times (Y_u)_{kj}^2 \times (C^{(1)}_{q u q d; jkfi})^2 \nonumber \\
	&& + G^{q u q d (1)}_{b} \times S_{b} \times \mathcal{G}^{q u q d (1)}_{b} \times (Y_u)_{kj}^2 \times (C^{(1)}_{q u q d; fkji})^2  \nonumber \\
	&& + G^{q u q d (8)}_{b} [singlet] \times S_{b} \times \mathcal{G}^{q u q d (8)}_{b} \nonumber \\
	&& \qquad \times (Y_u)_{kj}^2 \times (C^{(8)}_{q u q d; fkji})^2 \nonumber \\
	&& + G^{q d (1)}_{b} [singlet] \times S_{b} \times \mathcal{G}^{q d (1)}_{b} \times (Y_d^\dagger)_{kj}^2 \times (C^{(1)}_{q d; fkji})^2 \nonumber \\
	&& + G^{q d (8)}_{b} [singlet] \times S_{b} \times \mathcal{G}^{q d (8)}_{b} \times (Y_d^\dagger)_{kj}^2 \times (C^{(8)}_{q d; fkji})^2 \nonumber
\end{eqnarray}

\begin{eqnarray}
	&& \qquad\qquad (4 \pi)^2 \mu \frac{d}{d \mu} C^{(6)}_{q^2 d^2 H^2; fifi} = \\
	&& G^{q u q d (8)}_{a} \times S_{a} \times \mathcal{G}^{q u q d (8)}_{a} \times (Y_u)_{kj}^2 \times (C^{(8)}_{q u q d; jkfi})^2 \nonumber \\
	&& + G^{q u q d (8)}_{b} [octet] \times S_{b} \times \mathcal{G}^{q u q d (8)}_{b} \nonumber \\
	&& \qquad \times (Y_u)_{kj}^2 \times (C^{(8)}_{q u q d; fkji})^2 \nonumber \\
	&& + G^{q d (1)}_{b} [octet] \times S_{b} \times \mathcal{G}^{q d (1)}_{b} \times (Y_d^\dagger)_{kj}^2 \times (C^{(1)}_{q d; fkji})^2 \nonumber \\
	&& + G^{q d (8)}_{b} [octet] \times S_{b} \times \mathcal{G}^{q d (8)}_{b} \times (Y_d^\dagger)_{kj}^2 \times (C^{(8)}_{q d; fkji})^2 \nonumber
\end{eqnarray}

\begin{eqnarray}
	&& \qquad\qquad (4 \pi)^2 \mu \frac{d}{d \mu} C^{(5)}_{q^2 u^2 H^2; fifi} = \\
	&& G^{\ell e q u (1)}_{a} [quark] \times S_{a} \times \mathcal{G}^{\ell e q u (1)}_{a} \nonumber \\
    && \qquad \times (Y_e)_{kj}^2 \times (C^{(1)}_{\ell e q u; jkfi})^2 \nonumber \\
	&& + G^{q u q d (1)}_{a} \times S_{a} \times \mathcal{G}^{q u q d (1)}_{a} \times (Y_d)_{kj}^2 \times (C^{(1)}_{q u q d; fijk})^2 \nonumber \\
	&& + G^{q u q d (1)}_{b} \times S_{b} \times \mathcal{G}^{q u q d (1)}_{b} \times (Y_d)_{kj}^2 \times (C^{(1)}_{q u q d; jifk})^2 \nonumber \\
	&& + G^{q u q d (8)}_{b} [singlet] \times S_{b} \times \mathcal{G}^{q u q d (8)}_{b} \nonumber \\
	&& \qquad \times (Y_d)_{kj}^2 \times (C^{(8)}_{q u q d; jifk})^2 \nonumber \\
	&& + G^{q u (1)}_{b} [singlet] \times S_{b} \times \mathcal{G}^{q u (1)}_{b} \times (Y_u^\dagger)_{kj}^2 \times (C^{(1)}_{q u; fkji})^2 \nonumber \\
	&& + G^{q u (8)}_{b} [singlet] \times S_{b} \times \mathcal{G}^{q u (8)}_{b} \times (Y_u^\dagger)_{kj}^2 \times (C^{(8)}_{q u; fkji})^2 \nonumber
\end{eqnarray}

\begin{eqnarray}
	&& \qquad\qquad (4 \pi)^2 \mu \frac{d}{d \mu} C^{(6)}_{q^2 u^2 H^2; fifi} = \\
	&& G^{q u q d (8)}_{a} \times S_{a} \times \mathcal{G}^{q u q d (8)}_{a} \times (Y_d)_{kj}^2 \times (C^{(8)}_{q u q d; fijk})^2 \nonumber \\
	&& + G^{q u q d (8)}_{b} [octet] \times S_{b} \times \mathcal{G}^{q u q d (8)}_{b} \nonumber \\
    && \qquad \times (Y_d)_{kj}^2 \times (C^{(8)}_{q u q d; jifk})^2 \nonumber \\
	&& + G^{q u (1)}_{b} [octet] \times S_{b} \times \mathcal{G}^{q u (1)}_{b} \times (Y_u^\dagger)_{kj}^2 \times (C^{(1)}_{q u; fkji})^2 \nonumber \\
	&& + G^{q u (8)}_{b} [octet] \times S_{b} \times \mathcal{G}^{q u (8)}_{b} \times (Y_u^\dagger)_{kj}^2 \times (C^{(8)}_{q u; fkji})^2 \nonumber
\end{eqnarray}

\begin{eqnarray}
	&& \qquad\qquad (4 \pi)^2 \mu \frac{d}{d \mu} C^{(1)}_{\ell^4 H^2; fifi} = \\
	&& G^{\ell e}_{c} [\ell \ell] \times S_{c} \times \mathcal{G}^{\ell e}_{c} \times [Y_e Y_e^\dagger]_{jj} \times (C_{\ell e; fijj})^2 \nonumber \\
	&& + G^{\ell u}_{c} [lepton] \times S_{c} \times \mathcal{G}^{\ell u}_{c} \times [Y_u Y_u^\dagger]_{jj} \times (C_{\ell u; fijj})^2 \nonumber \\
	&& + G^{\ell d}_{c} [lepton] \times S_{c} \times \mathcal{G}^{\ell d}_{c} \times [Y_d Y_d^\dagger]_{jj} \times (C_{\ell d; fijj})^2 \nonumber \\
	&& + G^{\ell q (1)}_{c} [lepton] \times S_{c} \times \mathcal{G}^{\ell q (1)}_{c} \nonumber \\
	&& \qquad \times \left( [Y_u^\dagger Y_u]_{jj} + [Y_d^\dagger Y_d]_{jj} \right) \times (C^{(1)}_{\ell q; fijj})^2 \nonumber \\
	&& + G^{\ell q (3)}_{c} [lepton] \times S_{c} \times \mathcal{G}^{\ell q (3)}_{c} \nonumber \\
	&& \qquad \times \left( [Y_u^\dagger Y_u]_{jj} + [Y_d^\dagger Y_d]_{jj} \right) \times (C^{(3)}_{\ell q; fijj})^2 \nonumber \\
	&& + G^{\ell \ell}_{c} \times S_{c} \times \mathcal{G}^{\ell \ell}_{c} \times [Y_e^\dagger Y_e]_{jj} \times (C_{\ell \ell; fijj})^2 \nonumber \\
	&& + G^{\ell \ell}_{d} [singlet] \times S_{d} \times \mathcal{G}^{\ell \ell}_{d} \times [Y_e^\dagger Y_e]_{jj} \times (C_{\ell \ell; fjji})^2 \nonumber
\end{eqnarray}

\begin{eqnarray}
	&& \qquad\qquad (4 \pi)^2 \mu \frac{d}{d \mu} C^{(2)}_{\ell^4 H^2; fifi} = \\
	&& G^{\ell \ell}_{d} [triplet(2)] \times S_{d} \times \mathcal{G}^{\ell \ell}_{d} \times [Y_e^\dagger Y_e]_{jj} \times (C_{\ell \ell; fjji})^2 \nonumber
\end{eqnarray}

\begin{eqnarray}
	&& \qquad\qquad (4 \pi)^2 \mu \frac{d}{d \mu} C_{e^4 H^2; fifi} = \\
	&& G^{\ell e}_{c} [ee] \times S_{c} \times \mathcal{G}^{\ell e}_{c} \times [Y_e^\dagger Y_e]_{jj} \times (C_{\ell e; jjfi})^2 \nonumber \\
	&& + G^{q e}_{c} [lepton] \times S_{c} \times \mathcal{G}^{q e}_{c} \nonumber \\
	&& \qquad \times \left( [Y_u^\dagger Y_u]_{jj} + [Y_d^\dagger Y_d]_{jj} \right) \times (C_{q e; jjfi})^2 \nonumber \\
	&& + G^{e u}_{c} [lepton] \times S_{c} \times \mathcal{G}^{e u}_{c} \times [Y_u Y_u^\dagger]_{jj} \times (C_{e u; fijj})^2 \nonumber \\
	&& + G^{e d}_{c} [lepton] \times S_{c} \times \mathcal{G}^{e d}_{c} \times [Y_d Y_d^\dagger]_{jj} \times (C_{e d; fijj})^2 \nonumber \\
	&& + G^{e e}_{c} \times S_{c} \times \mathcal{G}^{e e}_{c} \times [Y_e Y_e^\dagger]_{jj} \times (C_{e e; fijj})^2 \nonumber \\
	&& + G^{e e}_{d} \times S_{d} \times \mathcal{G}^{e e}_{d} \times [Y_e Y_e^\dagger]_{jj} \times (C_{e e; fjji})^2 \nonumber
\end{eqnarray}

\begin{eqnarray}
	&& \qquad\qquad (4 \pi)^2 \mu \frac{d}{d \mu} C^{(1)}_{q^4 H^2; fifi} = \\
	&& G^{q e}_{c} [quark] \times S_{c} \times \mathcal{G}^{q e}_{c} \times [Y_e Y_e^\dagger]_{jj} \times (C_{q e; fijj})^2 \nonumber \\
	&& + G^{q u (1)}_{c} [qq] \times S_{c} \times \mathcal{G}^{q u (1)}_{c} \times [Y_u Y_u^\dagger]_{jj} \times (C^{(1)}_{q u; fijj})^2 \nonumber \\
	&& + G^{q d (1)}_{c} [qq] \times S_{c} \times \mathcal{G}^{q d (1)}_{c} \times [Y_d Y_d^\dagger]_{jj} \times (C^{(1)}_{q d; fijj})^2 \nonumber \\
	&& + G^{q u (8)}_{c} [qq] \times S_{c} \times \mathcal{G}^{q u (8)}_{c} \times [Y_u Y_u^\dagger]_{jj} \times (C^{(8)}_{q u; fijj})^2 \nonumber \\
	&& + G^{q d (8)}_{c} [qq] \times S_{c} \times \mathcal{G}^{q d (8)}_{c} \times [Y_d Y_d^\dagger]_{jj} \times (C^{(8)}_{q d; fijj})^2 \nonumber \\
	&& + G^{\ell q (1)}_{c} [quark] \times S_{c} \times \mathcal{G}^{\ell q (1)}_{c} \times [Y_e^\dagger Y_e]_{jj} \times (C^{(1)}_{\ell q; jjfi})^2 \nonumber \\
	&& + G^{\ell q (3)}_{c} [quark] \times S_{c} \times \mathcal{G}^{\ell q (3)}_{c} \times [Y_e^\dagger Y_e]_{jj} \times (C^{(3)}_{\ell q; jjfi})^2 \nonumber \\
	&& + G^{q q (1)}_{c} \times S_{c} \times \mathcal{G}^{q q (1)}_{c} \nonumber \\
	&& \qquad \times ( [Y_u^\dagger Y_u]_{jj} + [Y_d^\dagger Y_d]_{jj} ) \times (C^{(1)}_{q q; fijj})^2 \nonumber \\
	&& + G^{q q (3)}_{c} \times S_{c} \times \mathcal{G}^{q q (3)}_{c} \nonumber \\
	&& \qquad \times ( [Y_u^\dagger Y_u]_{jj} + [Y_d^\dagger Y_d]_{jj} ) \times (C^{(3)}_{q q; fijj})^2 \nonumber \\
	&& + G^{q q (1)}_{d} [singlet] \times S_{d} \times \mathcal{G}^{q q (1)}_{d} \nonumber \\
	&& \qquad \times ( [Y_u^\dagger Y_u]_{jj} + [Y_d^\dagger Y_d]_{jj} ) \times (C^{(1)}_{q q; fjji})^2 \nonumber \\
	&& + G^{q q (3)}_{d} [singlet] \times S_{d} \times \mathcal{G}^{q q (3)}_{d} \nonumber \\
	&& \qquad \times ( [Y_u^\dagger Y_u]_{jj} + [Y_d^\dagger Y_d]_{jj} ) \times (C^{(3)}_{q q; fjji})^2 \nonumber
\end{eqnarray}

\begin{eqnarray}
	&& \qquad\qquad (4 \pi)^2 \mu \frac{d}{d \mu} C^{(2)}_{q^4 H^2; fifi} = \\
	&& G^{q q (1)}_{d} [triplet(2)] \times S_{d} \times \mathcal{G}^{q q (1)}_{d} \nonumber \\
	&& \qquad \times ( [Y_d^\dagger Y_d]_{jj} - [Y_u^\dagger Y_u]_{jj} ) \times (C^{(1)}_{q q; fjji})^2 \nonumber \\
	&& + G^{q q (3)}_{d} [triplet(2)] \times S_{d} \times \mathcal{G}^{q q (3)}_{d} \nonumber \\
	&& \qquad \times ( [Y_u^\dagger Y_u]_{jj} - [Y_d^\dagger Y_d]_{jj} ) \times (C^{(3)}_{q q; fjji})^2 \nonumber
\end{eqnarray}


\begin{eqnarray}
	&& \qquad\qquad (4 \pi)^2 \mu \frac{d}{d \mu} C_{u^4 H^2; fifi} = \\
	&& G^{\ell u}_{c} [quark] \times S_{c} \times \mathcal{G}^{\ell u}_{c} \times [Y_e^\dagger Y_e]_{jj} \times (C_{\ell u; jjfi})^2 \nonumber \\
	&& + G^{q u (1)}_{c} [uu] \times S_{c} \times \mathcal{G}^{q u (1)}_{c} \nonumber \\
	&& \qquad \times ( [Y_u^\dagger Y_u]_{jj} + [Y_d^\dagger Y_d]_{jj} ) \times (C^{(1)}_{q u; jjfi})^2 \nonumber \\
	&& + G^{q u (8)}_{c} [uu] \times S_{c} \times \mathcal{G}^{q u (8)}_{c} \nonumber \\
	&& \qquad \times ( [Y_u^\dagger Y_u]_{jj} + [Y_d^\dagger Y_d]_{jj} ) \times (C^{(8)}_{q u; jjfi})^2 \nonumber \\
	&& + G^{e u}_{c} [quark] \times S_{c} \times \mathcal{G}^{e u}_{c} \times [Y_e Y_e^\dagger]_{jj} \times (C_{e u; jjfi})^2 \nonumber \\
	&& + G^{u d (1)}_{c} \times S_{c} \times \mathcal{G}^{u d (1)}_{c} \times [Y_d Y_d^\dagger]_{jj} \times (C^{(1)}_{u d; fijj})^2 \nonumber \\
	&& + G^{u d (8)}_{c} \times S_{c} \times \mathcal{G}^{u d (8)}_{c} \times [Y_d Y_d^\dagger]_{jj} \times (C^{(8)}_{u d; fijj})^2 \nonumber \\
	&& + G^{u u}_{c} \times S_{c} \times \mathcal{G}^{u u}_{c} \times [Y_u Y_u^\dagger]_{jj} \times (C_{u u; fijj})^2 \nonumber \\
	&& + G^{u u}_{d} \times S_{d} \times \mathcal{G}^{u u}_{d} \times [Y_u Y_u^\dagger]_{jj} \times (C_{u u; fjji})^2 \nonumber
\end{eqnarray}

\begin{eqnarray}
	&& \qquad\qquad (4 \pi)^2 \mu \frac{d}{d \mu} C_{d^4 H^2; fifi} = \\
	&& G^{\ell d}_{c} [quark] \times S_{c} \times \mathcal{G}^{\ell d}_{c} \times [Y_e^\dagger Y_e]_{jj} \times (C_{\ell d; jjfi})^2 \nonumber \\
	&& + G^{q d (1)}_{c} [dd] \times S_{c} \times \mathcal{G}^{q d (1)}_{c} \nonumber \\
	&& \qquad \times ( [Y_u^\dagger Y_u]_{jj} + [Y_d^\dagger Y_d]_{jj} ) \times (C^{(1)}_{q d; jjfi})^2 \nonumber \\
	&& + G^{q d (8)}_{c} [dd] \times S_{c} \times \mathcal{G}^{q d (8)}_{c} \nonumber \\
	&& \qquad \times ( [Y_u^\dagger Y_u]_{jj} + [Y_d^\dagger Y_d]_{jj} ) \times (C^{(8)}_{q d; jjfi})^2 \nonumber \\
	&& + G^{e d}_{c} [quark] \times S_{c} \times \mathcal{G}^{e d}_{c} \times [Y_e Y_e^\dagger]_{jj} \times (C_{e d; jjfi})^2 \nonumber \\
	&& + G^{u d (1)}_{c} \times S_{c} \times \mathcal{G}^{u d (1)}_{c} \times [Y_u Y_u^\dagger]_{jj} \times (C^{(1)}_{u d; jjfi})^2 \nonumber \\
	&& + G^{u d (8)}_{c} \times S_{c} \times \mathcal{G}^{u d (8)}_{c} \times [Y_u Y_u^\dagger]_{jj} \times (C^{(8)}_{u d; jjfi})^2 \nonumber \\
	&& + G^{d d}_{c} \times S_{c} \times \mathcal{G}^{d d}_{c} \times [Y_d Y_d^\dagger]_{jj} \times (C_{d d; fijj})^2 \nonumber \\
	&& + G^{d d}_{d} \times S_{d} \times \mathcal{G}^{d d}_{d} \times [Y_d Y_d^\dagger]_{jj} \times (C_{d d; fjji})^2 \nonumber
\end{eqnarray}


Other dimension-8 operators not shown in Tab.~\ref{tab:dimension_8} (see Ref.~\cite{Murphy:2020rsh} for their complete list) are not considered in the renormalization of double-insertions; see discussion in Sec.~\ref{sec:large_Yukawa_contributions}.

Note that double-insertions of $ Q^{(3)}_{\ell e q u} $ turn out to be finite.
(This is the only case found for which the divergence in double-insertions at intermediate stages contains a term proportional to the Levi-Civita symbol.)
This contradicts Ref.~\cite{Cirigliano:2017tqn}, that claims a divergence and exploits the corresponding leading logarithm to constrain the Wilson coefficient of $ Q^{(3)}_{\ell e q u} $. 


In the RG equations above: $ \mathcal{G} $ correspond to the residue of the $ \epsilon $-pole extracted from the calculation indicated in Eq.~\eqref{eq:identifying_div} times the overall factor seen in Eq.~\eqref{eq:definition_gamma0} (without symmetry and group factors), $ S $ designate symmetry factors, and $ G $ group factors (subscripts $ a, b, c, d $ designate different topologies). They are given as follows:


\begin{eqnarray}
	&& \mathcal{G}^{\ell e d q}_{a} = \mathcal{G}^{\ell e q u (1)}_{a} = \mathcal{G}^{q u q d (1)}_{a} = \mathcal{G}^{q u q d (8)}_{a} = -4 \,, \nonumber\\ 
	&& \mathcal{G}^{q u q d (1)}_{b} = \mathcal{G}^{q u q d (8)}_{b} = 2 \,, \nonumber\\ 
	&& \mathcal{G}^{\ell e}_{c} = \mathcal{G}^{q x (1)}_{c} = \mathcal{G}^{q x (8)}_{c} = \mathcal{G}^{e e}_{c} = \mathcal{G}^{x x}_{c} \nonumber \\
	&& = \mathcal{G}^{\ell \ell}_{c} = \mathcal{G}^{q q (1)}_{c} = \mathcal{G}^{q q (3)}_{c} = \mathcal{G}^{\ell q (1)}_{c} = \mathcal{G}^{\ell q (3)}_{c} \nonumber \\
	&& = \mathcal{G}^{e x}_{c} = \mathcal{G}^{\ell x}_{c} = \mathcal{G}^{q e}_{c} = \mathcal{G}^{u d (1)}_{c} = \mathcal{G}^{u d (8)}_{c} = 2 \,, \nonumber\\ 
	&& \mathcal{G}^{\ell e}_{b} = \mathcal{G}^{q x (1)}_{b} = \mathcal{G}^{q x (8)}_{b} = -16 \,, \nonumber\\ 
	&& \mathcal{G}^{\ell \ell}_{d} = \mathcal{G}^{q q (1)}_{d} = \mathcal{G}^{q q (3)}_{d} = \mathcal{G}^{e e}_{d} = \mathcal{G}^{x x}_{d} = 2 \,, 
\end{eqnarray}


\begin{eqnarray}
	&& S_{b} = \frac{1}{2} \,, \quad S_{a} = \frac{1}{2} \,, \quad S_{d} = 1 \,, \quad S_{c} = 1 \,,
\end{eqnarray}



\begin{eqnarray}
	&& G^{\ell \ell}_{d} [singlet] = \frac{1}{2} \,, \; G^{\ell \ell}_{d} [triplet(2)] = \frac{1}{2} \,, \; G^{\ell \ell}_{c} = 1 \,, \nonumber \\
	&& G^{q q (1)}_{d} [singlet] = \frac{1}{2} \,, \; G^{q q (1)}_{d} [triplet(2)] = \frac{1}{2} \,, \nonumber \\
	&& G^{q q (1)}_{c} = N_c \,, \nonumber \\
	&& G^{q q (3)}_{d} [singlet] = \frac{5}{2} \,, \; G^{q q (3)}_{d} [triplet(2)] = \frac{3}{2} \,, \nonumber \\
	&& G^{q q (3)}_{c} = N_c \,, \nonumber \\
	&& G^{\ell q (1)}_{c} [lepton] = N_c \,, \; G^{\ell q (1)}_{c} [quark] = 1 \,, \nonumber \\
	&& G^{\ell q (3)}_{c} [lepton] = N_c \,, \; G^{\ell q (3)}_{c} [quark] = 1 \,, \nonumber \\
	&& G^{e e}_{d} = 1 \,, \; G^{e e}_{c} = 1 \,, \nonumber \\
	&& G^{x x}_{d} = 1 \,, \; G^{x x}_{c} = N_c \,, \nonumber \\
	&& G^{e x}_{c} [lepton] = N_c \,, \; G^{e x}_{c} [quark] = 1 \,, \nonumber \\
	&& G^{u d (1)}_{c} = N_c \,, \nonumber \\
	&& G^{u d (8)}_{c} = \frac{1}{4} \left( 1 - \frac{1}{N_c} \right) \,, \nonumber \\
	&& G^{\ell e}_{c} [\ell \ell] = 1 \,, \; G^{\ell e}_{c} [e e] = 1 \,, \nonumber \\
	&& G^{\ell x}_{c} [lepton] = N_c \,, \; G^{\ell x}_{c} [quark] = 1 \,, \nonumber \\
	&& G^{q e}_{c} [lepton] = N_c \,, \; G^{q e}_{c} [quark] = 1 \,, \nonumber \\
	&& G^{q x (1)}_{c} [q q] = N_c \,, \; G^{q x (1)}_{c} [x x] = N_c \,, \nonumber \\
	&& G^{q x (8)}_{c} [q q] = \frac{1}{4} \left( 1 - \frac{1}{N_c} \right) \,, \nonumber \\
	&& G^{q x (8)}_{c} [x x] = \frac{1}{4} \left( 1 - \frac{1}{N_c} \right) \,,
\end{eqnarray}
and

\begin{eqnarray}
	&& G^{\ell e}_{b} = 1 \,, \\
	&& G^{q x (1)}_{b} [singlet] = \frac{1}{N_c} \,, \; G^{q x (1)}_{b} [octet] = 2 \,, \nonumber \\ 
	&& G^{q x (8)}_{b} [singlet] = \frac{1}{4} \left( N_c - \frac{2}{N_c} + \frac{1}{N_c^3} \right) \,, \nonumber \\
	&& G^{q x (8)}_{b} [octet] = \frac{1}{2} \frac{1}{N_c^2} \,, \nonumber \\ 
	&& G^{\ell e d q}_{a} [lepton] = N_c \,, \; G^{\ell e d q}_{a} [quark] = 1 \,, \nonumber \\
	&& G^{q u q d (1)}_{b} = 1 \,, \; G^{q u q d (1)}_{a} = N_c \,, \nonumber \\
	&& G^{q u q d (8)}_{b} [singlet] = \frac{1}{4} \left( 1 - \frac{1}{N_c^2} \right) \,, \nonumber \\
	&& G^{q u q d (8)}_{b} [octet] = \frac{1}{2} \left( N_c - \frac{2}{N_c} \right) \,, \; G^{q u q d (8)}_{a} = \frac{1}{2} \,, \nonumber \\ 
	&& G^{\ell e q u (1)}_{a} [lepton] = N_c \,, \; G^{\ell e q u (1)}_{a} [quark] = 1 \,. \nonumber
\end{eqnarray}


Everywhere in this paper
$T^A$ are the $SU(3)$ generators, normalized such that $\text{tr} \{ T^A T^B \} = \frac{1}{2} \delta^{A B}$ (a different normalization was taken in Ref.~\cite{ValeSilva:2022vsl}).
Also, $ \tau^I $ are the Pauli matrices, for which $ \text{tr} \{ \tau^I \tau^J \} = 2 \delta^{I J} $.

As one cross-check, doing an independent calculation we verify the SM \cite{Herrlich:1993yv,Herrlich:1996vf} at the leading order.
For the cancellation of logarithmic terms in case (II) of the SM (see Sec.~\ref{sec:effective_operators}), note that we have two possible internal flavours, up- and charm-quarks, which is beyond the scope of the previous expressions, where double-insertions of the same operator with the same flavour content have been considered. However, one can easily depict the cancellation in the following way: since all interactions are left-handed, the only relevant diagram is the bottom right one in Fig.~\ref{fig:diagrams_2}; there are two contributions from the diagram with two internal charm-quark lines, and one contribution from each of the two diagrams with internal up- and charm-quark lines, where the lines to which the two scalars couple (representing now insertions of vacuum expectation values) are the charm-quark ones; the latter contributions carry the opposite sign with respect to the former ones, which follows from the use of the unitarity of the CKM matrix, so the total result vanishes in the end. In case (III) of the SM, below the EW scale only the latter diagrams, carrying internal up- and charm-quark lines, are present, so there is no (super-hard) GIM cancellation in this case. For completeness, case (I) of the SM only involves internal up-quarks below the EW scale.


\section{Sensitivity of meson mixing to New Physics effects}\label{sec:sensitivity_NP}

Bounds on the size of NP in the neutral meson systems $ K $ and $ B_{(s)} $ are discussed in Refs.~\cite{Charles:2013aka, Charles:2020dfl} (see also Refs.~\cite{UTfit:2007eik,Descotes-Genon:2018foz}). There, since these observables are used in the global extraction of the elements of the CKM matrix in the SM (the mass differences $ \Delta m_{d,s} $ playing presently a more relevant role in the fit than the indirect CP violating quantity $ | \epsilon_K | $), the extraction of the elements of the CKM matrix is redone allowing for NP contamination in the fit:
NP is parametrized under the form

\begin{equation}
    M_{12}^i = ( M_{12}^i )_\text{SM} \times (1 + h_i \times e^{2 i \sigma_i}) \,,
\end{equation}
and combined bounds on $ h_i $ and $ \sigma_i $ are extracted, where the index $ i $ refers to the different neutral meson systems. Ref.~\cite{Charles:2013aka} combines contributions from NP in the kaon system with the top-up (case (I) discussed in Sect.~\ref{sec:effective_operators}) set of contributions from the SM, keeping cases (II) and (III) unmodified, which is consistent with our discussion in Sect.~\ref{sec:phenomenology_tops_loops}.
The extracted ranges are $ h_d < 0.26 $, $ h_s < 0.12 $ \cite{Charles:2020dfl},
and $ | h_K \times \sin (2 \sigma_K + 2 {\rm Arg} (V_{td} V_{ts}^\ast)) | \lesssim 0.6 $ \cite{Charles:2013aka};
all bounds are $ 95 $\% Confidence Level intervals.
These bounds constrain NP at the energy scale relevant for the different observables, namely, $ \sim M_{B_{(s)}} $ for the $ B_{(s)} $ mass differences, and $ 2 $~GeV for the indirect CP violating quantity in the kaon sector. To constrain different kinds of NP at the EW scale, one introduces the short-distance QCD corrections collected in Ref.~\cite{Buras:2001ra}, that provide the running and mixing of $ | \Delta F | = 2 $ four-fermion contact interactions below the EW scale at the Next-to-Leading Order (NLO). One also needs the relevant bag parameters, which are taken from Ref.~\cite{Carrasco:2015pra} in the $ K $ system, and
Ref.~\cite{Dowdall:2019bea}
for the $ B_{(s)} $ systems (both for $ N_f = 2 + 1 + 1 $). Light quark masses are taken from Ref.~\cite{FlavourLatticeAveragingGroup:2019iem} ($ N_f = 2 + 1 + 1 $), see also references therein. Put together, bounds on NP at the scale $ \mu_t = m_t (m_t) = 166 $~GeV follow the following pattern for the Wilson coefficients $ C_1 {(i)}, C_2 {(i)}, C_3 {(i)}, C_4 {(i)}, C_5 {(i)} $ introduced in the main text, see Eq.~\eqref{eq:low_energy_basis}:


\begin{alignat}{6}
%
	&\Delta m_{d,s} \, &&: \quad 1 &&\,:\, \phantom{0} 2 &&\,:\, \phantom{0} 0.4 &&\,:\, \phantom{00} 5 &&\,:\, \phantom{0} 2 \,, \\ 
	&| \epsilon_K | \, &&: \quad 1 &&\,:\, 40 &&\,:\, 10 &&\,:\, 100 &&\,:\, 30 \nonumber
\end{alignat}
with respect to bounds on the Wilson coefficient $ C_1 {(i)} $ (for instance, the bound on $ | C_4 {(B_{(s)})} | $ is about $ 5 $ times stronger than the bound on $ | C_1 {(B_{(s)})} | $); bounds on the latter are

\begin{alignat}{2}
	& |C_1 {(B)}| &&< (1 \times 10^3 \; \text{TeV})^{-2} \,, \\
	& |C_1 {(B_s)}| &&< (3 \times 10^2 \; \text{TeV})^{-2} \,, \nonumber\\
	& | \text{Im} \{ C_1 {(K)} \} | &&< (2 \times 10^4 \; \text{TeV})^{-2} \,. \nonumber 
\end{alignat}
The running effects between the NP scale $ \Lambda_{\rm NP} $ and the EW scale $ \mu_{EW} = \mu_t $ are the concern of the main text.
Although short-distance QCD effects below the EW scale are being included for NP, we are not including QCD effects above the EW scale, which are suppressed by a relatively small strong coupling.

{A final comment is in order: in kaon meson mixing we employ the value of $\kappa_\epsilon$ encoding non-local effects of up-quarks calculated in Ref.~\cite{Lenz:2010gu}, compatible with Ref.~\cite{Buras:2010pza}, and assume that possible NP contamination therein is small; an exploratory Lattice QCD study of this non-perturbative effect has been made in Ref.~\cite{Christ:2015phf}. The possibility of NP in direct CP violation in the kaon system is discussed in the main text.}


\bibliography{mybib}{}
\bibliographystyle{unsrturl}

\end{document}